\documentclass{article}
\usepackage{algorithmic}
\newcommand{\cij}{c_{i,j}}
\newcommand{\Cij}{C_{i,j}}
\newcommand{\dsijm}{\Delta_{i,j}^m}
\newcommand{\dijm}{\dsijm(\cij)}
\newcommand{\dsijr}{\Delta_{i,j}^r}
\newcommand{\dijr}{\dsijr(\cij)}
\newcommand{\dsijn}{\Delta_{i,j}^n}
\newcommand{\dijn}{\dsijn(\cij)}
\newcommand{\dsim}{\Delta_{i}^m}
\newcommand{\dlim}{\dsim(f_i)}
\newcommand{\dsir}{\Delta_{i}^r}
\newcommand{\dlir}{\dsir(f_i)}
\newcommand{\dsin}{\Delta_{i}^n}
\newcommand{\dlin}{\dsin(f_i)}
\newtheorem{remark}{Remark}[section]
\newtheorem{Theorem}[remark]{Theorem}
\newtheorem{Definition}[remark]{Definition}
\begin{document}
\title{Improving the LP bound of a MILP by branching concurrently}
\author{H. Georg B{\"u}sching\\
        H\"andelstra{\ss}e 9\\
        32457 Porta Westfalica\\
        Germany}
\maketitle 

\begin{abstract}
In this paper the branching trees for attacking MILP are reviewed. 
Under certain circumstances branches can be done concurrently.
This is fully investigated with the result that there are restrictions for certain dual values and reduced costs.
As a sideeffect of this study a new class of cuts for MILP is found, which are defined by those values.
\end{abstract}
\section{Motivation of the following thoughts}
Nowadays the technique for doing MILP (Mixed Integer Linear Programming) is based on the branch and bound method. This method uses the best solution of the linear inequality system with objective function (= LP-instance) by leaving out the integer conditions from the mixed integer linear inequality system with objective function (= MILP-instance). Then this method searches for an $0$-$1$ (or integer) variable $x_n$, which has a non-integer value $q_n$. The next step is to create two new LP-instances by adding first $x_n = 0$ (or $x_n \le [q_n]$) and secondly $x_n = 1$ ($x_n \ge [q_n + 1]$). By continuing this process a binary tree of problems is created.
\\
 Now take two different nodes in this tree, so you look at two different LP-instances. With both problems it is possible that some $x_n$ is still not integer. We'll create a branch on that variable for both problems. It can happen at a big and sparse MILP-instance, that the same similar branching will lead to exactly the same calculations at the new LP-instances. From a numerical point of view this is unsatisfactory.
\\
\\
New ideas have been developed here which use some kind of independence of branching. These will help to prevent such double calculations. One further aim of these new techniques is a better measurement and control of what happens at a branching. A practical and short-term outcome should be better limits for huge MILP-instances. It should be noticed that the prominent group of huge Traveling Salesman Problems is a part of this group. As a matter of fact, this group was indeed the starting point of the author's thoughts about this topic.
\\
\\
We'll show that the combination of branches can be described by an ordinary linear inequality system, so that the problem to get an optimal combination of branches will be a LP-instance (luckily not a MILP-instance). We'll reach this formulation at the middle of the second section at theorem \ref{CentralPF1}. We want to use instead of a binary tree of depth $n$, which has $2^n$ problems, just $2n$ problems. We'll try to combine the solutions of the $2n$ problems as well as possible to get a bound for the original problem (MILP-instance), which will be better than the LP-bound but normally not as good as the bound by solving all $2^n$ problems.
\\
\\
We'll furthermore see that it is even possible to define a very huge LP for each MILP, which represents the ability to combine the several case differentiations. 
\\
\\
The main idea is not too difficult: \\
\\
We'll measure the differences of the dual variables and the gain of the objective function when creating new problems, which each has one inequality more than the starting LP-instance. These differences of the dual variables are naturally connected to the branches. 
Then we'll choose those differences of dual variables, so that for all combinations of choices at the connected branches, all dual inequalities will hold for sure. 
By adding the gain of each chosen branching, we get a total gain, which gives a better limit of the original problem. \\
\\
\\
It should be noted that the whole paper has been fully elaborated by the author. 
In fact the only real reference are the basic facts about LPs as presented in \cite{LP}.
\section{Description of the technique in a very broad context}
\subsection{Basic terminology and central theorem}
In the following we examine a problem $P$, which can be partially represented as a minimal linear problem $P_l$. The desription of the whole problem needs some additional case differentiations. It should be remarked that the set of MILP-instances is a real subset of this problem class. The linear problem $P_l$ has a set $\{y_m\}$ of inequalities in variables $\{x_n\}$. Without loss of the generality we assume that all $\{y_m\}$ are $\ge$-inequalities. We'll argue later why equations may be excluded from the scope. Furthermore we expect that the inequalities named by $y_m$ represent all bounds to the $x_n$.
\\
Since the term branching has been used in LP-terminology in quite a lot of places with an emphasize of really creating of one problem two problems, a new terminology will be introduced. We use the terms of cases and files instead. A case $\Cij$ will stand for the evaluation of one possibility $j$ of a case differentiation $i$, the sum of the cases $\Cij$ make together a file $F_i$, which will be in other words the case differentiation. But we'll soon define it more concretely. We shall examine, when and how the files can be combined to get a higher lower limit for the optimal solution.\\
\\
Therefore we'll start from the dual point of view, so $\omega$ will be considered as the objective function of the dual problem. To ease the notification, we state that the indices $\{m\}$ and $\{n\}$ have empty intersection. Via defining the index $\{r\}$ as the union of both we get something that will help us in making all formulation much easier.\\
\\
The dual solution space of $P_l$ will be noted as $V_0$ with optimal subspace $L_0$ and optimal value $\omega_0$. Furthermore we chose an arbitrary $y \in L_0$. By looking at one case j of the case differentiation $i$ the dual solution space is $V_{i,j}$. But we'll restrict this solution space by $ \{ \omega_{i,j} \ge \omega_0 \} $ to get a polytope $P_{i,j}$. Now let the case $\Cij$ be the following set $\{\cij |\ y + \cij \in P_{i,j}\}$. So it is a movement to 0 of $P_{i,j}$. Our objective function $\omega_{i,j}$ can easily expanded to $\Cij$ just by setting it to $\omega_{i,j}(\cij)$, which is the same as $\omega_{i,j}(y + \cij) - \omega_{i,j}(y)$ due to the linearness of $\omega_{i,j}$.\\
\begin{Definition}
\label{CentralDefinition}
Now choose $\cij \in \Cij$ and define $\dijm = y_m - (y + \cij)_m = -(\cij)_m$ as the change of the $m$-th dual variable. Also define $\dijn = r_{n}(y) - r_{x}(y+\cij)$, where $r_{n}$ shall be the reduced cost of $x_n$. By this we also define $\dsijr$.
\end{Definition}

It is important to state that the $\dijn$ can be calculated by the $\dijm$ with additional info of the value of the new dual variable(s) $y+{\cij}$. If we have equalities as conditions, we'll see that those dual values give no interesting $\Delta^m$-values, but the $\Delta^n$-values can be calculated with the help of these values. We also define the vector $l = (y, r_n(y))$, this vector has coordinates in {r}.\\
As $y+\cij$ is a dual solution, all coordinates $y_m$ most be positive or null. The same holds for the linear function $r_{n}$, the value $R_{x_n}(y+\cij)$ must be positive or null. Putting these facts together, we get the following remark, which already has the structure of our main statement \ref{CentralAB}:
\begin{remark}
$$ \dijr  \le  l_r $$
\label{FirstIQ}
\end{remark}

For later purpose we also investigate the property of linearity of the $\dijm$ and the $\dijn$ and so of the $\dijr$. We easily see that everything is linear:
\begin{remark}
\label{FirstLin}
$$
\begin{array}{lll}
\dsijm(\lambda\cij)&=& \lambda\dijm\\
\dsijm(\cij + d_{i,j})&=&\dijm + \dsijm(d_{i,j})\\
\dsijn(\lambda\cij)&=&\lambda\dijn\\
\dsijn(\cij + d_{i,j})&=&\dijn + \dsijn(d_{i,j})\\
\end{array}
$$
\end{remark}
In \ref{FirstIQ} we call those inequalities, where the right hand side is greater $0$ the main inequalities.
\begin{remark}
If $\cij$ relates to an optimal solution of $P_{i,j}$, then one main inequality is sharp.  
\label{mainIQ}
\end{remark}
Too see this we assume that this is not the case. We consider $y + (1 + \epsilon) \cij$, which has higher objective value and the inequalities in \ref{FirstIQ} still hold. As those inequalities make the dual variable related to the inequalities of $P_{i,j}$ positive and the dual inequalities of $P_{i,j}$ true. We follow that $(1 + \epsilon) \cij$ is in $\Cij$. This is a contradiction to the optimality of $\cij$. So one of the main inequalities must be sharp. The other non-main inequalities are in fact trivial, since the values here for $y$ itself are already sharp, so the $\Delta$-values must be negative or null.
\\

The next step of our thoughts is to go from a case to the case differentiations, which will be named as files as announced.
Let $F_i = \bigoplus_j\Cij$ be a file. If we take an element $f_i = \oplus_j \cij$ out of our construct $F_i$. We further define:

$$ \omega_i(f_i) = \min_j (\omega_{i,j}(\cij)) $$
$$ \dlir = \max_j (\dijr) $$
\\
The delta represent the highest differences of the changes within a case differentiation (file) of the dual variables and the dual inequalities.\\
Based on \ref{FirstLin} you get easily the following equalities and inequalities.
\begin{remark}
\label{SecondLin}
$$
\begin{array}{lll}
\dsir (\lambda f_i)&=&\lambda \dlir\\
\dsir (f_i + g_i)&\le&\dlir + \dsir (g_i)\\
\omega_i(\lambda f_i)&=&\lambda \omega_i(f_i)\\
\omega_i(f_i + g_i)&\ge&\omega_i(f_i) + \omega_i(g_i)\\
\end{array}
$$
\end{remark}
So you can conclude that the deltas are still convex and $\omega_i$ is concave.\\
Now we even build a more complex space $\cal{P}$, which will be the sum of all files and our final object. This space represents parallel files.\\
$$ {\cal{P}} = \bigoplus_i F_i$$
with the following functions for $p \in \cal{P}$:
$$\Delta^r_{\cal{P}}(p) = \sum_i \dlir$$
$$\omega_{\cal{P}}(p) = \omega_0 + \sum_i \omega_i(f_i)$$

Now let $p \in \cal{P}$, then we have chosen in all cases of all files a solution vector. Remember, that we are always talking about dual solutions and variables. If we chose for each file $F_i$ a case $C_{i,j(i)}$, then we have a new problem $\hat P_l$, which is in fact $P_l$ together with the inequalities from all $C_{i,j(i)}$. For this we can calculate a solution $\hat y$ by the means of the $\Delta$: If we look at $\hat y_m$, then the value is: 
$$\hat y_m = y_m - (\sum_i \Delta^m_{i,j(i)}(c_{i,j(i)})) \ge y_m - \sum_i \dlim = y_m - \Delta^m_{\cal{P}} (p)$$
So by $\Delta^m_{\cal{P}} (p) \le y_m$, we can make sure, that the dual variable $\hat y_m$ is positive. Since this is not important for equations, we only considered inequalities before. Notice also that the condition is independent of our choice $j(i)$.\\
\\
For the validity of the $n$-th dual inequality we got something similar:
$$
\begin{array}{llll}
\Delta^n_{\cal{P}}(p)&\le&r_{n}(y)&\Rightarrow\\
\sum_i \Delta^n_{i,j(i)} (c_{i, j(i)})&\le&r_{n}(y)&\Leftrightarrow\\
0&\le&r_{n}(y)\\
\end{array}
$$
Keep in mind, that also $r_{n}(\hat y)$ can be calculated by the $\Delta^m_{i,j(i)}(c_{i,j(i)})$ with the help of new dual variables of all chosen cases. This is the same as the $\dijn$ could be derived from the $\dijm$, as we have seen before. \\
\\
The objective value of $\hat y$ is $\omega_0 + \sum_i \omega_i (c_{i,j(i)}) \ge \omega_{\cal{P}}(p)$. Putting these thoughts together we get our central statement.
\begin{Theorem}[Central theorem]
\label{CentralAB}
If for all $r$ holds, that $\Delta^r_{\cal{P}}(p) \le l_r$, then the original problem $P$ must have an optimal solution that is greater than $\omega_{\cal{P}}(p)$, which is greater than the original $\omega_0$ of the linear problem $P_l.$
\end{Theorem}
\subsection{Building the little combining LP}
The last statement seems to be rather abstract, but by an easy trick, we'll get two different forms, that can be used in an algorithm. To get the first we just substitute $f_i$ by $\lambda_i f_i$ with $\lambda_i \ge 0$. As the deltas and the objectives $\omega_i$ are linear on a scalar (\ref{SecondLin}), we get the main result of this article:
\begin{Theorem}[Central theorem - simple form]
The ability to combine case differentiations can be assured by the following inequalities:
$$ \sum_i \lambda_i \dlir \le l_r $$
The new lower limit is $\sum_i \lambda_i \omega_i(f_i) + \omega_0$. So by solving this LP-instance in $\lambda_i$ we get better lower limit for our problem $P$.
\label{CentralPF1}
\end{Theorem}
By looking at all $P_{i,j}$ all values in this LP-instance can be calculated, first the $(y + \cij)_m$ and $R_{x_n}(y + \cij)$, secondly $\dijm$, $\dijn$, $\omega(c_{i,j})$ and lastly the $\dlim$, $\dlin$ and the $\omega_i(f_i)$. \\
\\
By the definition of $\omega_i$ it is natural to choose the $\cij$ in such a way that for all $j$ the equation $\omega_i(f_i) = \omega_{i,j}(\cij)$ holds. This can be achieved by substitution of $\cij$ by $\omega_i(f_i) \omega(\cij)^{-1}\cij$. This is for our purposes well-defined, because when $\omega(\cij) = 0$ then we get no progress on the objective function of $P$ from this case differentiation. By this substitution in \ref{CentralPF1}, the objective function $\omega_i(f_i)$ remains the same, but normally the $\Delta$-values will decrease, leading to higher values when using the practical form of the central theorem. We call this trick \emph{normalization}.\\
\\
As a next step we want to generalize \ref{CentralPF1}. We substitute in \ref{CentralAB} $f_i$ by $\sum \lambda_{i,k} f_{i,k}$ and use the convexness of $\Delta_i^r$ (\ref{SecondLin}) to get the second sum in the upcoming theorem \ref{CentralPF2}.\\
Suppose furthermore that you have not only got one optimal solution of $P_{LP}$, but another solutions $y + c_0 \in L_0$. This other solution can be found in a case $C_0 = F_0$, which by itself already will be a file $F_0$. We have furthermore a natural function $\Delta^r$ defined as above on this case. In the argumentation to \ref{CentralAB} we could have introduced this special case without any problems. By this we get an extra term of $- c_0$ in the calculation of $\hat y_m$. Since $y + c_0$ is an  optimal solution of $P_l$, there will be no quality growth directly related to $c_0$. So we don't have to define a function $\omega$ for it. \\
\\
\begin{Theorem}[Central theorem - more complex form]
The ability to combine case differentiations can be assured by the following inequalities:
$$ \sum_k \lambda_{0,k} \Delta^r(c_0^k) + \sum_{i,k} \lambda_{i,k} \Delta_i^r(f_i^k) \le l_r $$
The new lower limit is $\sum_{i,k} \lambda_{i,k} \omega_i(f_i^k) + \omega_0$. So by solving this LP-instance $S$ in $\lambda_{i,k}$ and $\lambda_{0,k}$ we get a better lower limit for our problem $P$.
\label{CentralPF2}
\end{Theorem}
Although this formulation seems to be much stronger than the version 1, this is not really the case. Looking at the second sum we see, that the $\lambda_{i,k}$ can be created in \ref{CentralPF1} by choosing $F_{i_1,j} = F_{i_2,j}$ for $i_1 \ne i_2$. This is possible because it was never stated that we made a case differentiation only once. \\
But by the creation of \ref{CentralPF2} we see something different: If some $\lambda_{i,k_1}$ and $\lambda_{i,k_2}$ with $k_1 \ne k_2$ are non-null for an optimal solution of the resulting LP-instance in \ref{CentralPF2}, then we can find better values by setting $f_{i,\hat k} = \lambda_{i,k_1} f_{i,k_1} + \lambda_{i,k_2} f_{i,k_2}$. So by generating new columns we can sometimes improve the bound for $P$.\\
Also the first sum in \ref{CentralPF2} has some limitations. Suppose the following values: $y_m = 1$, $(y + c_0^k)_m = 0$, $\Delta_{i_1}^m (f_{i_1}) = \Delta_{i_2}^m (f_{i_2}) = 1$. Then the files for $i_1$ and $i_2$ cannot be combined fully. But if you exchange $y$ and $y + c_0^k$, they can, because then $\Delta_{i_1}^m (f_{i_1}) = \Delta_{i_2}^m (f_{i_2}) = 0$ holds. So we should not expect that the first sum in \ref{CentralPF2} to help us very much for our needs. \\
\\
We have presented in this section a theory on a problem, which is described by a LP too weak. 
But in truth we studied the dual LP, where the problem is described too sharp and can be weakened by case differentiations. 
As \ref{CentralPF1} can be weakened by natural case differentiation by fixing for one $i$ and one $j$ $\dsir = \dsijr$, we could use the whole theory on it. 
This self-appliance is surprising and fascinating. 
We will sketch one manual example later. 
Even if this looks interesting on the first glance, not much progress on the lower bound is expected by this iteration.
\\
\\
Although the mathematical formulation to combine case differentiations (files) has been explained broadly in this section, some details are still not covered. The problem is that those details might be not too easy to attack at all. When you think of a fast implementation of this idea you want to have an effective, numerical stable and fast algorithm to find good elements $\cij \in C_{i,j}$, where most $\cij$ are zero. By these you get good $f_i$ for a given solution $y$. We'll see later in \ref{Implement} that the normal approach to use optimal solution of the $P_{i,j}$ leads in some examples to problems. 
So later in \ref{Theory} and \ref{Practise} we will attack these problems by using non-optimal $y + \cij$ even before normalization, where most $\dijm$ and $\dijn$ should be zero.

\subsection{Building the huge combining LP}
In this section we will follow again the made definitions and results and reach a mathematical satisfactory formulation of the theory.\\
We had started with one solution $y \in V_0$. For each case of a file we have also a $y_{i,j} \in V_{i,j}$. 
Notice that the reduced costs of one variable of $y_0$ (and $y_{i,j}$) are by definition just a linear equation dependent of the $(y_0)_n$ ($(y_{i,j})_n$). 
So we define as in \ref{CentralDefinition} the $\dsijr$ as variables which are calculated by linear equations from $y_0$ and $y_{i,j}$, the same holds for the $\omega_{i,j}$, which are also linear dependant on $y_0$ and $y_{i,j}$.
\\
Via the restrictions $\dsir \ge \dsijr$ and $\omega_i \le \omega_{i,j}$ for all $j$ we have defined $\dsir$ and $\omega_i$ as linear inequalities. 
Like before we define $\Delta^r_{\cal{P}}$ as a sum of the $\dsir$ and $\omega_{\cal{P}}$ as the sum of the $\omega_i$ plus the objective $\omega$ of $P_l$, which is also a linear term of $y_0$. 
Via using the restrictions of \ref{CentralAB} and setting the objective to $\omega_{\cal{P}}$ we have defined now a very huge LP $P_{\mbox{comb}}$. 
We can now easily formulate a theorem, which describes the problem of doing case differentiations in parallel in a mathematical satisfactory way:
\begin{Theorem}[Central theorem - complete form]
\label{BigCombiningLP}
Each solution of $P_{\mbox{comb}}$ represents a lower bound of $P$. 
\end{Theorem}

The optimal value is the optimal lower bound possible via our combining technique. \\
We restricted ourselves from writing all inequalities explicitely down, as the huge amount of indices for each variable might only be confusing and all inequalities have already been described impliciteley.

But it should be noticed that the definitions in \ref{CentralPF1}, \ref{CentralPF2} and \ref{DualLPForCut}
 are just tightened and less complex inequalities systems than this system. \\
Consider that $P$ has $n$ variables and $m$ restrictions and that all variables are binaries, so a simple case differentation can be made on each variably, then this new LP would have at least $2nm$ variables. For small sized problems this might still numerical possible to be calculated. \\
This huge LP should not be attacked for optimal values in the author's view because of the not avoidable high computing time but searched for good solutions in a effective manner.\\
It should be mentioned that it is possible to use the theory again and to formulate a construction a LP of dimension $4n^2m$ which should give better lower bounds then the $P_{\mbox{comb}}$. \\

\section{Implementation with usage of optimal solutions of the subproblems \label{Implement}}
Putting the thoughts from the previous section together you get the following algorithm described as pseudo-Code to get higher objective values of a MILP-instance with only $0$-$1$ variables:\\\\

\begin{algorithmic} [1]
\STATE derive $P_l$ from given MILP-instance
\STATE load $P_l$ into LP-solver
\STATE solve $P_l$ and save one optimal solution $x$ and the fitting dual solution $y$
\FOR{all $i$, where $(x)_i$ is not integer}
\STATE {case $j = 1$:}
\STATE add inequality $x_i \le 0$ to $P_l$ (so getting $P_{i,1}$)
\STATE solve this LP by usage of the old dual solution $y$ and get $y + c_{i,1}$ as dual solution \label{first}
\STATE calculate all $\Delta_{i,1}^r(c_{i,1})$- \COMMENT {as defined in \ref{CentralDefinition}} 
\STATE {case $j = 2$:}
\STATE add inequality $x_i \ge 1$ to $P_l$ (so getting $P_{i,2}$)
\STATE solve this LP by usage of the old dual solution $y$ and get $y + c_{i,2}$ as dual solution \label{second}
\STATE calculate all $\Delta_{i,2}^r(c_{i,2})$ - \COMMENT {as defined in \ref{CentralDefinition}}
\STATE $\omega_i(f_i) = \min (\omega_{i,1}(c_{i,1}), \omega_{i,2}(c_{i,2}))$  
\IF{$\omega_i(f_i) > 0$}
\STATE mark $i$
\IF{Normalization trick is wanted} 
\IF{$\omega_{i,1}(c_{i,1}) \le \omega_{i,2}(c_{i,2})$}
\STATE for all $r$: $\Delta_{i,2}^r(c_{i,2}) = \omega_{i,1}(c_{i,1}) (\omega_{i,2}(c_{i,2}))^{-1} \Delta_{i,2}^r(c_{i,2})$ 
\ELSE
\STATE for all $r$: $\Delta_{i,1}^r(c_{i,1}) = \omega_{i,2}(c_{i,2})\*(\omega_{i,1}(c_{i,1}))^{-1}\Delta_{i,1}^r(c_{i,1})$ 
\ENDIF 
\ENDIF
\STATE for all $r$: $\dlim = \max (\Delta_{i,1}^r(c_{i,1}),\Delta_{i,2}^r(c_{i,2}))$
\ENDIF
\ENDFOR
\STATE build new LP-instance $R$ with all marked $i$ in variables $\lambda_i$ as described in \ref{CentralPF1}
\STATE solve $R$
\end{algorithmic}

In first implementation I was not able to use the old solution in lines \ref{first} and \ref{second} effectively. This has quite some impact because of the degeneration of the optimal dual solution in most of the prominent problems. \\
To understand this let's consider you have chosen an optimal $y$ with $(y)_m > 0$ for a problem $P$. But also an optimal $\bar y$ exists with $(\bar y)_m = 0$. Now for all $i$ one $\cij$ could exist where $(y + \cij)_m = 0$. This leads to the situation that no file could be combined ensured by the inequality of $R$, which deals with the fact that dual variables for inequalities should be positive. But if you had chosen the other $\bar y$, you would have less problems. As a side-remark it should be noticed that also all $P_{i,j}$ normally then have a degenerate dual solution space. \\
The way to use the old solution $y$ as a starting point for solving $P_{i,j}$ has two benefits: First the optimal solution should be found faster numerically and secondly normally when dealing with degeneracy the above described effect should happen less often.\\
The above algorithm has been implemented with the Open-Source package glpk. In this program all MILP are transformed to be Minimum-problems by exchanging the sign of the objective. So some results on MILP have an extraordinary sign.  The problem library MIPLIB2003 has been processed partially getting the results on the following page.\\
\\
In the given table the column \emph{Branches} measures the number of variables where branching took place. The actual number of calculated LPs is 2 times more plus the initials LP and the combining LP. The column \emph{Degree} is equal to $\sum_i \lambda_i$ of the optimal value of the combining LP. It gives an idea how much branches can be used at the same time but also in an effective way. The current implementation separates already the normal and the dual LP because of future plans. So we measure both in seconds. Furthermore also the total time for all calculation is presented. \\
The given table only includes those instances, where the program finished within 1 hour. Furthermore for some instances no advantage at all was made, because at no branch there was an increase in both nodes at all. For more investigations these problems might be put out of scope. On the other hand for the instance tr12-30 a quite high lower bound was reached: Starting from 14210 the bound 79695 is reached which is much nearer to the real value at 130596.\\

\small
\begin{tabular}{l|rrrrrrr}
Instance&Pure LP&Bound Inc&Branches&Degree&Normal&Dual&Total
\\
\hline
10teams & 917.00 & 0.00 & 159 & 0.00 & 3 & 1 & 11\\
a1c1s1 & 997.53 & 1195.33 & 173 & 53.92 & 5 & 1 & 24\\
aflow30a & 983.17 & 14.28 & 31 & 5.19 & 1 & 0 & 2\\
aflow40b & 1005.66 & 7.16 & 38 & 1.80 & 3 & 2 & 17\\
air04 & 55535.44 & 84.61 & 292 & 1.00 & 59 & 21 & 1040\\
air05 & 25877.61 & 72.54 & 223 & 1.00 & 28 & 5 & 329\\
arki001 & 7579599.81 & 126.65 & 81 & 6.63 & 5 & 2 & 12\\
cap6000 & -2451537.33 & 0.00 & 2 & 0.00 & 12 & 5 & 18\\
danoint & 62.64 & 0.05 & 34 & 1.00 & 0 & 1 & 3\\
disctom & -5000.00 & 0.00 & 251 & 0.00 & 69 & 0 & 129\\
fiber & 156082.52 & 15734.31 & 47 & 6.90 & 1 & 0 & 3\\
fixnet6 & 1200.88 & 210.51 & 60 & 21.33 & 1 & 0 & 2\\
gesa2 & 25476489.68 & 81043.25 & 58 & 35.91 & 2 & 1 & 5\\
gesa2-o & 25476489.68 & 81891.56 & 73 & 36.70 & 1 & 0 & 4\\
glass4 & 800002400.00 & 0.00 & 72 & 0.00 & 1 & 0 & 1\\
harp2 & -74353341.50 & 0.00 & 30 & 0.00 & 3 & 1 & 6\\
liu & 346.00 & 214.00 & 536 & 1.00 & 2 & 1 & 16\\
manna81 & -13297.00 & 0.00 & 872 & 0.00 & 10 & 1 & 92\\
markshare1 & 0.00 & 0.00 & 6 & 0.00 & 0 & 0 & 0\\
markshare2 & 0.00 & 0.00 & 7 & 0.00 & 0 & 0 & 0\\
mas74 & 10482.80 & 42.52 & 12 & 1.19 & 1 & 0 & 1\\
mas76 & 38893.90 & 24.86 & 11 & 1.62 & 0 & 0 & 0\\
misc07 & 1415.00 & 0.00 & 31 & 0.00 & 1 & 0 & 1\\
mkc & -611.85 & 0.00 & 105 & 0.00 & 5 & 0 & 16\\
mod011 & -62121982.55 & 0.00 & 16 & 0.00 & 13 & 1 & 16\\
modglob & 20430947.62 & 69955.22 & 29 & 8.31 & 0 & 0 & 0\\
mzzv11 & -22945.24 & 0.00 & 836 & 0.00 & 68 & 1 & 323\\
mzzv42z & -21623.00 & 0.00 & 676 & 0.00 & 48 & 1 & 278\\
net12 & 17.25 & 11.40 & 429 & 1.30 & 27 & 89 & 3115\\
noswot & -43.00 & 0.00 & 28 & 0.00 & 1 & 0 & 1\\
nsrand-ipx & 48880.00 & 0.00 & 67 & 0.00 & 37 & 2 & 61\\
opt1217 & -20.02 & 0.00 & 29 & 0.00 & 1 & 0 & 1\\
p2756 & 2688.75 & 10.20 & 30 & 2.00 & 3 & 0 & 4\\
pk1 & 0.00 & 0.00 & 15 & 0.00 & 0 & 0 & 0\\
pp08a & 2748.35 & 762.82 & 51 & 11.41 & 0 & 0 & 0\\
pp08aCUTS & 5480.61 & 166.85 & 46 & 6.47 & 1 & 0 & 1\\
protfold & -41.96 & 0.00 & 449 & 0.00 & 7 & 1 & 34\\
qiu & -931.64 & 0.00 & 36 & 0.00 & 1 & 1 & 3\\
roll3000 & 11097.13 & 5.44 & 214 & 4.32 & 6 & 1 & 36\\
rout & 981.86 & 2.34 & 35 & 1.00 & 0 & 1 & 1\\
set1ch & 32007.73 & 3904.90 & 138 & 64.56 & 1 & 0 & 2\\
seymour & 403.85 & 1.50 & 632 & 3.30 & 30 & 4 & 291\\
sp97ar & 652560391.11 & 241502.97 & 194 & 2.00 & 89 & 12 & 522\\
swath & 334.50 & 0.40 & 45 & 5.71 & 5 & 1 & 19\\
timtab1 & 28694.00 & 137970.93 & 136 & 16.46 & 0 & 0 & 1\\
timtab2 & 83592.00 & 106311.17 & 233 & 27.02 & 0 & 1 & 4\\
tr12-30 & 14210.43 & 65484.48 & 348 & 322.01 & 1 & 0 & 8\\
vpm2 & 9.89 & 0.48 & 31 & 7.41 & 1 & 0 & 1\\

\end{tabular}
\normalsize

\section{Two manual examples of the presented technology}
\subsection{A trivial one}

$$y_{1,1}: x_{1,1} + x_{1,2} \ge 1 \mbox{ and }y_{1,2}: x_{1,2} + x_{1,3} \ge 1 \mbox{ and }y_{1,3}: x_{1,3} + x_{1,1} \ge 1$$
$$y_{2,1}: x_{2,1} + x_{2,2} \ge 1 \mbox{ and }y_{2,2}: x_{2,2} + x_{2,3} \ge 1 \mbox{ and }y_{2,3}: x_{2,3} + x_{2,1} \ge 1$$

Minimize $\omega(x) = x_{1,1} + x_{1,2} + x_{1,3} + x_{2,1} + x_{2,2} + x_{2,3}$ and all variables have to be integer.\\

Clearly by just viewing the problem the optimal value of $\omega$ is $4$. The optimal solution of the LP itself is $x_{i,l} = 0.5$ for $i\in\{1;2\}$ and $l\in\{1;2;3\}$ with objective $3$. All dual variables $y_{i,l}$ have also the value $0.5$. The next step is to make the case differentiations. Let's concentrate on $x_{i,1}$. Either $x_{i,1} = 0$ holds (case $j = 1$) or $x_{i,1} \ge 1$ (case $j = 2$). \\

Calculating the 4 different LPs we get the following values for the dual variables and objectives:

$$
\begin{array}{cccccc}
i=1, j=1:& y_{1,1} = 1,& y_{1,2} = 0,& y_{1,3} = 1,& y_{2,l} = 0.5,& \omega = 3.5\\
i=1, j=2:& y_{1,1} = 0,& y_{1,2} = 1,& y_{1,3} = 0,& y_{2,l} = 0.5,& \omega = 3.5\\
i=2, j=1:& y_{2,1} = 1,& y_{2,2} = 0,& y_{2,3} = 1,& y_{1,l} = 0.5,& \omega = 3.5\\
i=2, j=2:& y_{2,1} = 0,& y_{2,2} = 1,& y_{2,3} = 0,& y_{1,l} = 0.5,& \omega = 3.5\\
\end{array}
$$

The reduced costs of the $x_{i,l}$ are not of interest because all variables had in the LP-version of the problem no reduced costs.

Following the definitions of the preceding main chapter we get for the $\dijm$ the following values:

$$
\begin{array}{clllcc}
i=1, j=1:& \Delta_{1,1}^{1,1} = -0.5,& \Delta_{1,1}^{1,2} = 0.5,& \Delta_{1,1}^{1,3} = -0.5,& \Delta_{1,1}^{2,l} = 0,& \omega_{1,1} = 0.5\\
i=1, j=2:& \Delta_{1,2}^{1,1} = 0.5,& \Delta_{1,2}^{1,2} = -0.5,& \Delta_{1,2}^{1,3} = 0.5,& \Delta_{1,2}^{2,l} = 0,& \omega_{1,2} = 0.5\\
i=2, j=1:& \Delta_{2,1}^{2,1} = -0.5,& \Delta_{2,1}^{2,2} = 0.5,& \Delta_{2,1}^{2,3} = -0.5,& \Delta_{2,1}^{1,l} = 0,& \omega_{2,1} = 0.5\\
i=2, j=2:& \Delta_{2,2}^{2,1} = 0.5,& \Delta_{2,2}^{2,2} = -0.5,& \Delta_{2,2}^{2,3} = 0.5,& \Delta_{2,2}^{1,l} = 0,& \omega_{2,2} = 0.5\\
\end{array}
$$

This gives the files values:
$$
\begin{array}{cccccc}
i=1:& \Delta_{1}^{1,1} = 0.5,& \Delta_{1}^{1,2} = 0.5,& \Delta_{1}^{1,3} = 0.5,& \Delta_{1}^{2,l} = 0,& \omega_{1} = 0.5\\
i=2:& \Delta_{2}^{2,1} = 0.5,& \Delta_{2}^{2,2} = 0.5,& \Delta_{2}^{2,3} = 0.5,& \Delta_{2}^{1,l} = 0,& \omega_{1} = 0.5\\
\end{array}
$$
So we reach the following LP for the combination of the files.

$$
\begin{array}{c}
\mbox{Validity that }y_{1,1} \ge 0: 0.5\lambda_1 + 0\lambda_2 \le 0.5\\ 
\mbox{Validity that }y_{1,2} \ge 0: 0.5\lambda_1 + 0\lambda_2 \le 0.5\\ 
\mbox{Validity that }y_{1,3} \ge 0: 0.5\lambda_1 + 0\lambda_2 \le 0.5\\ 
\mbox{Validity that }y_{2,1} \ge 0: 0\lambda_1 + 0.5\lambda_2 \le 0.5\\ 
\mbox{Validity that }y_{2,2} \ge 0: 0\lambda_1 + 0.5\lambda_2 \le 0.5\\ 
\mbox{Validity that }y_{2,3} \ge 0: 0\lambda_1 + 0.5\lambda_2 \le 0.5
\end{array}
$$ 

With the following objective $0.5 \lambda_1 + 0.5 \lambda_1$. As the objective of this problem is 1 we can derive that the lower limit of the original MILP is at least $3 + 1 = 4$. As there are solutions with this objective, this limit is sharp\\

This trivial example also gives the right idea that for case differentiations in different part of LP, which are not connected, that the files can be combined. \\

\subsection{Almost a real one}
\label{Almost a real one}
$$
\begin{array}{c}
x_{1} + x_{2} + x_{3} \le 1 \\ 
x_{2} + x_{3} + x_{4} \le 1 \\
x_{3} + x_{4} + x_{5} \le 1 \\
x_{4} + x_{5} + x_{1} \le 1 \\
x_{5} + x_{1} + x_{2} \le 1
\end{array}
$$

Maximize $\omega(x) = x_{1} + x_{2} + x_{3} + x_{4} + x_{5}$ and all variables have to be integer.

Let $i\in\{1;\ldots;5\}$, the optimal dual solution of the LP is simply $y_i = \frac{1}{3}$ with $\omega = \frac{5}{3}$. 
The optimal  solution of the MILP has objective of $1$.
For example we branch on the two cases $x_1=0$ and $x_1\ge1$. We get then the following dual solutions:

$$
\begin{array}{ccllllll}
x_1=0:& y =& (0& 0.5& 0.5& 0& 0.5)& \mbox{ with } \omega = 1.5\\
x_1=1:& y =& (1& 0& 0& 1& 0)& \mbox{ with } \omega = 1\\
x_1=1:& y =& (0.5& 0.25& 0.25& 0.5& 0.25)& \mbox{ with } \omega = 1.5 \mbox{(Normalization!)}
\end{array} 
 $$
 Naturally the theory can also be applied to maximum problems. So we get for $\Delta_1$:
 $$ \Delta_1 = (\frac{1}{3} \frac{1}{12} \frac{1}{12} \frac{1}{3} \frac{1}{12}) \mbox{ with } \omega = \frac{1}{6}$$
 Via using the symmetry of the problem we get the combination of the files the following LP:
 
 $$
 \begin{array}{rrrrrrrrrrr}
 \frac{1}{3}\lambda_1&  +& \frac{1}{12}\lambda_2& +&\frac{1}{3}\lambda_3& +&\frac{1}{12}\lambda_4& +& \frac{1}{12}\lambda_5& \le& \frac{1}{3} \\
 \frac{1}{12}\lambda_1&  +& \frac{1}{3}\lambda_2& +&\frac{1}{12}\lambda_3& +&\frac{1}{3}\lambda_4& +& \frac{1}{12}\lambda_5& \le& \frac{1}{3} \\
 \frac{1}{12}\lambda_1&  +& \frac{1}{12}\lambda_2& +&\frac{1}{3}\lambda_3& +&\frac{1}{12}\lambda_4& +& \frac{1}{3}\lambda_5& \le& \frac{1}{3} \\
 \frac{1}{3}\lambda_1&  +& \frac{1}{12}\lambda_2& +&\frac{1}{12}\lambda_3& +&\frac{1}{3}\lambda_4& +& \frac{1}{12}\lambda_5& \le& \frac{1}{3} \\
 \frac{1}{12}\lambda_1&  +& \frac{1}{3}\lambda_2& +&\frac{1}{12}\lambda_3& +&\frac{1}{12}\lambda_4& +& \frac{1}{3}\lambda_5& \le& \frac{1}{3} \\  
 \end{array}
 $$
 
 With objective $\frac{1}{6}\lambda_1 + \frac{1}{6}\lambda_2 + \frac{1}{6}\lambda_3 + \frac{1}{6}\lambda_4 + \frac{1}{6}\lambda_5$.\\
 The optimal solution is $\lambda_i = \frac{4}{11}$ and objective is $\frac{10}{33}$. So that we have shown that the maximum in our original MILP is less or equal $\frac{5}{3} - \frac{10}{33}$.
 
 At this point it is again possible to make a case differentiation on $x_1 = 0$ or $x_1 \ge 1$. If we assume $x_1 = 0$ the above LP would have the following form:
 
 $$
 \begin{array}{rrrrrrrrrrr}
 \frac{1}{3}\lambda_1&  +& \frac{1}{12}\lambda_2& +&\frac{1}{3}\lambda_3& +&\frac{1}{12}\lambda_4& +& \frac{1}{12}\lambda_5& \le& \frac{1}{3} \\
 -\frac{1}{6}\lambda_1&  +& \frac{1}{3}\lambda_2& +&\frac{1}{12}\lambda_3& +&\frac{1}{3}\lambda_4& +& \frac{1}{12}\lambda_5& \le& \frac{1}{3} \\
 -\frac{1}{6}\lambda_1&  +& \frac{1}{12}\lambda_2& +&\frac{1}{3}\lambda_3& +&\frac{1}{12}\lambda_4& +& \frac{1}{3}\lambda_5& \le& \frac{1}{3} \\
 \frac{1}{3}\lambda_1&  +& \frac{1}{12}\lambda_2& +&\frac{1}{12}\lambda_3& +&\frac{1}{3}\lambda_4& +& \frac{1}{12}\lambda_5& \le& \frac{1}{3} \\
 -\frac{1}{6}\lambda_1&  +& \frac{1}{3}\lambda_2& +&\frac{1}{12}\lambda_3& +&\frac{1}{12}\lambda_4& +& \frac{1}{3}\lambda_5& \le& \frac{1}{3} \\  
 \end{array}
 $$
 
 And for $x_1 \ge 1$:
 
 $$
 \begin{array}{rrrrrrrrrrrr}
 -\frac{2}{3}\lambda_1&  +& \frac{1}{12}\lambda_2& +&\frac{1}{3}\lambda_3& +&\frac{1}{12}\lambda_4& +& \frac{1}{12}\lambda_5& \le& \frac{1}{3} \\
 \frac{1}{3}\lambda_1&  +& \frac{1}{3}\lambda_2& +&\frac{1}{12}\lambda_3& +&\frac{1}{3}\lambda_4& +& \frac{1}{12}\lambda_5& \le& \frac{1}{3} \\
 -\frac{2}{3}\lambda_1&  +& \frac{1}{12}\lambda_2& +&\frac{1}{3}\lambda_3& +&\frac{1}{12}\lambda_4& +& \frac{1}{3}\lambda_5& \le& \frac{1}{3} \\
 \frac{1}{3}\lambda_1&  +& \frac{1}{12}\lambda_2& +&\frac{1}{12}\lambda_3& +&\frac{1}{3}\lambda_4& +& \frac{1}{12}\lambda_5& \le& \frac{1}{3} \\
 -\frac{2}{3}\lambda_1&  +& \frac{1}{3}\lambda_2& +&\frac{1}{12}\lambda_3& +&\frac{1}{12}\lambda_4& +& \frac{1}{3}\lambda_5& \le& \frac{1}{3} \\  
 \end{array}
 $$
 
 We'll stop the calculation at this point. We could now calculate some $\Delta_1$ via checking the differences for the resulting normal variables $\lambda_i$ and build a new LP, 
 which would represent the possibility of combining the changes of $\lambda_i$ when doing the case differentiations. By this we would again reach better upper limit for the original problem.\\
 
 It possible to iterate this method until infinity, but some manual calculations have shown that the real lower limit will never be reached in this way.\\
 
 The author likes this example pretty much. It shows that non-trivial combining are possible, and that the method can be iterated in a surprising way. But it also shows some limits. 
 The above MILP is easily solved by doing the 4 case differentiations on $x_1$ and $x_2$. Furthermore it is even possible to make another case differentiation on one inequality.
 It is clear via the first inequality that $x_1 = 1$ or $x_2 = 1$ or $x_3 = 1$ or $x_1 = x_2 = x_3 = 0$ holds. Via this case differentiation it is seen most quickly seen that the optimal value 
 of the MILP is $1$. The author thinks that such case differentiation on inequalities should be investigated as an alternative to the normal branching on one variable 
 especially in the 0-1 MILP-context.\\
 
 \section{Effectiveness for finding good dual values in the branching LPs}
 \subsection{Sidestep: Searching for integrity}
 Only loosely connected to the rest of the paper we now investigate those MILPs and the derived LPs which have non-degenerate optimal solution space.
 As for all branching investigations especially in this paper the number of non-integers variables, which are supposed to be integer, should be as little as possible to reduce the running time of an implementation.
 \\
 Therefor we assume that we have an optimal solution vector $x_0$.
 We just freeze the objective function to the optimal value, so getting an additional equality.
 We set additional bounds on all integer variables $x_n$ via $[(x_0)_n] \le x_n \le [(x_0)_n] + 1$. 
 This is a good valid definition also for MILPs which are not binary problems. 
 In general it might be useful to try out use some bounds like $[(x_0)_n - \epsilon] \le x_n \le [(x_0)_n - \epsilon] + 1$.
 This gives some integer variables more freedom to become non-integer to allow other integer to more integer in the general MILP case.
 \\
 Finally we now define for all integer variables $n$ the objective of the minimization problem to enhance the variables to become integer.
 $$ c_n =  1 - 2([(x_0)_n] - (x_0)_n)$$
 The new vector $x_1$ is now got by solving this LP to optimality. 
 We now calculate again new $c_n$ in the described manner so that we have an iterative process.
 With this definition we have a good tool which gives almost integers a good motivation to become integer not hindering others in this process to give up integrity.
 \\
 \\
 Notice that the choice of the $c_n$ was done by experiments.
 It cannot be reasoned yet, why this choice was in the experiment superior to other approaches.
 Also the convergence of the method has only investigated by experiments.
 It is imaginable that reducing the number of non-integers might enhance the quality of some heuristic cuts findings, but the author has not received in his limited experiments any valuable result.
 For sure for the class of this paper in chapter \ref{cuts generation} this is not relevant as the cuts are only defined by certain dual values and reduced costs.
 \\
 \\
 Clearly this presented idea was motivated by the feasibility pump \cite{FeasPump} to generate integer solutions. 
 We present it also here because the following method was developed in spirit of this easy algorithm.
 
 \subsection{Measurement of good dual values}
 \label{Theory}
 We will again concentrate on \ref{CentralPF1}. 
 Looking at the inequality there you see that each file eats up certain inequalities (dual values) or variables (reduced costs).
 So to find good values, you have to search for files and hereby for dual variables who eat less of our stock but still give a good improvement in the objective function.
 First we have to define what it is the meaning is of eating up the stock of inequalities and variables.
 \\
 Suppose again you have made a case $j$ of case differentiation $i$ with a better dual variable set. 
 Some of the some the $\dsijr$ might be negative, but when the file is glued together by maximizing we suspect that the value $\dsir$ will be positive. 
 Anyhow even if it is really negative, quite likely no other case differentiation will need the negativeness.
 So we have argued to measure all negative $\dsijr$ as 0. 
 As a general approach measure distance to the starting point $y_0$ we can now define:
 
 $$D = \sum_r d_r \max(0, \dsijr)$$
 
 We also needed $D \ge 0$ to make the below algorithm work.
 \\
 \\
 We leave out the problem of setting the $d_r$ values, but first use this definition to get better dual values.
 Therefor we create more artificial variables $z_r$ in the dual space. 
 Via $z \ge 0$ and $z \ge \dsijr$ and $D = \sum_r z_r$ we reflect the definition.
 The idea is now to subtract $D$ from the objective $\omega$ in that way the optimal value of the LP created by the case $j$ of the case differentiation will have the same objective in our new artificial LP as $y_0$. 
 Let $D^0_{i,j}$ be the difference of this solution and $\omega_{i,j}$ the increase of the objective.
 Then we set:
 $$\omega^{\mbox{adj}}_{i,j} = \omega - \frac{D \omega_{i,j}}{D^0_{i,j}}$$ 
 So we have found an objective with the desired property.
 \\
 \\
 Via our definitions we have assured that $D$ is always positive. 
 When the new LP is now solved to optimality and point $y^{\mbox{new}}_{i,j}$ is found, it is therefor clear that its optimal value is in the polytope $P_{i,j}$. 
 Furthermore the following can easily be proved:
 $$ \frac{\omega^{\mbox{new}}_{i,j} - \omega_0}{D^{\mbox{new}}_{i,j}} \ge  \frac{\omega_{i,j} - \omega_0}{D_{i,j}} $$
 So in terms of effiency of eating up the stock the new point is better or equal than the first optimal point. 
 When it is equal, then the space $P_{i,j}$ might be often one dimensional. 
 But additionally by our definition we have not given up the wish for good objective gain.
 \\
 \\
 Also this trick can easily be iterated, visible already by our definitions.
 \\
 The algorithm can easily be enhanced that it works on finding better values of files, but this generalizations will not be presented here. 
 Also in an implementation you could use the values of the already manipulated cases of the case differentiation. 
 When an inequality or a variable has been eaten up a bit the new case should have this meal for free. 
 \\
 \\
 We now have had some fun with preparing effecient meals of inequalities and variables, but one crucial point of the receipt is still open: the definition of the $d_r$.
 If you define all $d_r = 1$, then the big $l_r$ in \ref{CentralPF1} will get two much attention.
 Tiny $l_r$, which might always hinder the combining of the files, are overlooked. \\
 So the natural choice is to set $d_r = \frac{1}{l_r}$ which will give all non-null inequalities and all variables with real reduced costs the same weight.
 Sadly this theoretically good approach lead in the author implementations to numerical problems. 
 Often the manipulated LP was bad conditioned.
 So the author suggests to use a lower limit like $0.01$ for all $d_r$.
 \\
 \\
 The author has implemented the above algorithm partially, but with some disappointment for him. 
 He didn't manage to use the old optimal solution in the software package glpk, so each manipulated LP had to be solved from scratch.
 This lead to too long running times.
 He thinks also that this time increase is only partial because of some missing features of the used software. 
 Using the theoretical good reasoned approach of this section might just be too numerical complex because of the sheer number of added constraints.
  
 \subsection{Finding quickly the dual values}
 \label{Practise}
 In the preceding subsection we described a theory to find dual solutions with good objective which could be considered as near to the basic dual solution $y_0$.
 We did this via introducing variables, which measured the distance to the original.
 Another approach in finding good dual solutions and so files is to use additional inequalities.
 Depending on the aim this can result in files which fit better to each other or in dual solution which can be calculated very quickly.
 \\
 Suppose you have already made a branching with a file $f_1$. Then to combine a second branching with the first you just demand:
 $$ l_r - \Delta_{1,j}^r \ge \Delta_{2,j}^r \forall r, j$$
 Speaking in terms of dual inequalities you get lower bounds for those variables $y_m$, which were nonzero in the basic dual solution.
 Furthermore restrictive dual inequalities which weren't in the solution vector $y_0$ become in general more restrictive. 
 The benefit of this approach is that using those additional restrictions it is clear that the two branching can fully combined.
 In terms of \ref{CentralPF1} this means that $\lambda_1 = \lambda_2 = 1$. \\
 Naturally the idea can easily be iterated via demanding:
 
 $$ l_r - \sum_i \Delta_{i,j}^r \ge \Delta_{i+1,j}^r \forall r, j$$

 Let's do at this point another sidestep. 
 Suppose that both files consist of two cases, which is the normal case for MILPs. 
 You have done the case differentiations to combine these two cases, so you have solved $2 + 2 = 4$ calculations.
 But doing instead a case differentiation on the the 4 cases ($2 * 2 = 4$), which already enumerate all possible combinations, you would have the same calculation time.
 But you will have a least better objective increase with these 4 cases than with combining the two case differentation.
 So for a clever implementation of these sketched algorithms the principle of parallel branching should not followed too strictly.
 Doing all case differentiations on $p$ cases might be interesting, when $2^p$ ist still comparable to $2p$.
 \\
 \\
But let's get back to additional restrictions for the dual inequalities.
We start with a metaphor: Linear equations describe the nature. 
When a butterfly flies up in Brazil the emerging circulations won't normally influence the weather in Europe.
Speaking in terms of LP an introduction of a new variable in a dual inequality has often only effect in those inequalities, which are strongly bound to the related inequalities.
So it is striking thought a neighborhood of a new variable, and to freeze all other variables, which are not in the neighborhood.
This should have a big reduction of the running time as a result.
If we have good criteria for the neighborhood the objective increase will often be comparable to the objective increase of the new dual inequality without freezing.
So it is quite likely that the resulting files might also be effective in terms of the last subsection.
\\
\\
Clearly defining neighborhood by the graph of the inequality system or other means is a complex story. 
The definition of the neighborhood should also be dependent on the type of the MILP. 
\\
Suppose you have a good neighborhood definition.
Then the technique of freezing most of the dual variables might also be an alternative to the strong branching method, which determines in a branch and cat framework the next variable to branch on.
The strong branching relies on a good and steep implementation of the dual simplex, where you use the values of the objective after only some iteration of the Simplex algorithm.
It should be noticed that such a steep dual Simplex algorithm is not a prerequisite of the algorithm.
So my approach can be used in less advanced packages like glpk to do something similar.
\\
\\
This chapter could be described as visionary or even dreamy, anyhow the subject of this paper is to present the author's idea on the subject.
To make it complete the author had just add it, otherwise he would always think that his idea have not been presented decently.
 
\section{Application of the theory to produce cuts for the original MILP}
\label{cuts generation}
When thinking of building in the look ahead term of the concurrent branching into a existing branch and cut framework, the dual combining inequality of \ref{CentralPF1} doesn't fit easily.
It is striking that instead of that additional dual LP you would like just to have more restrictions in normal space instead.
Generation of cuts should be the aim.
We'll see in this end of this chapter that this is possible.
\\
\\ 
We start at that point that we have an optimal dual solution $y^0$ with only one file $f_1$, which describes a case differentiation.
We now use a new special form of \ref{BigCombiningLP}, we freeze the $f_i$ as linear factors of a scalar $\lambda$. 
Contrary to the special form \ref{CentralPF1} we let $y$ really play the role of dual variables and not fix it to $l_r$-values. 
So we have as variables the vector $y$ and the scalar $\lambda$.
Transformed back via dual-dual correspondence this will give us more or less the normal inequalities and equalities plus one additional equality, which will be our cut.
But let's stick to the details.
We have the following dual inequalities for this special model:

\begin{Definition}
\label{DualLPForCut}
$$ x^{\mbox{add}}_m: \lambda \dsim(f_i) \le y_m \forall m $$
$$ x_n: \sum_m c_{m, n}y_k + \lambda \dsin(f_i) \le c_n \forall n $$
\end{Definition}

Where $(n)$ goes over all normal variables, $(m)$ over all dual variables (inequalities + equalities), $c_{m, n}$ are the matrix coeffizient of the LP and $c_n$ are the coeffizients of the normal objective.
The objective function of this dual problem is $\sum_m b_m y_m + \lambda \omega(f_i)$, where the $b_m$ are the right hand side of the inequalities and equalities.
\\
\\
This dual problem can be transformed to normal space:

\begin{remark}
\label{AlmostNewCut}
$$ y_m: \sum_n c_{m, n} x_n - x^{\mbox{add}}_m \ge b_m \forall m$$
$$ \lambda: \sum_m \dsim(f_i) x^{\mbox{add}}_m + \sum_n \dsin(f_i) x_n \ge \omega(f_i)$$
\end{remark}

The objective is just $\sum_n c_n x_n$ as the normal objective.
This looks already interesting, but prior the final transformation to get a cut we must first proof that this system is valid for all integer values.
Sadly the proof is very indirect, a direct proof was not discovered by the author.
\\
\\
Before doing the proof we must first study the reuse of files for other basic solutions than the starting one.
In \ref{CentralPF1} we had some files, which were tried to be added to some basic solution. 
If we would have used another solution with other $l_r$-values, we can naturally use the methodology also.
Adding the files might still be possible.
The only thing which might happen that all $\lambda_i$ in \ref{CentralPF1} have to be $0$. 
Same holds if we have a more strict LP. 
Then the dual solution has only some more variables, but the original ones are still there.
\begin{remark}
\label{UsefullnessForBranchAndCut}
The lookup term via combining files can still be used to a more strict version of the starting normal LP. 
For incompatible problems it can only be defined, when the missing dual variables (m) of the LP, where the file should be applied, the $\dsim$ are negative or $0$. 
\end{remark}

This remark also clarifies the usage of the lookup terms for integration in branch and cut frameworks.
The file info of a LP remains valid for all descendants and is normally invalid for other descendants of the root LP.
\\
Consider you have an integer solution of the LP.
Then it is clear that this integer solution is the only optimal solution of a version of the LP, which has been made more strict via adding more inequalities.
This more restrictive LP is represented in the dual space by a loosened LP. 
We still can try to add our file in the dual space. 
Via this we get the special model \ref{DualLPForCut} for the loosened dual inequality.
For this model the optimal objective has to be identical to the dual LP and the normal LP.
Otherwise we would prove that the optimal integer solution of the more strict LP has to have bigger objective than the already existing integer solution, which is a contradiction.
\\
The optimal dual solution of our loosened LP \ref{DualLPForCut} in the dual space is a solution of a more restrictive LP \ref{AlmostNewCut} than the original one.
So we have found an optimal normal solution, which also holds for the more restrictive inequalities.
As we had said that the original solution was the only optimal solution, it most be identical to the new one. 
So the original solution has to fulfill \ref{AlmostNewCut}.
\\
So all integer solutions fulfill it.
\\
\\
As a final step we can state that the normal inequality system is equivalent to:

$$ y_m: \sum_n c_{m, n} x_n \ge b_m \forall m$$
$$ \lambda: \sum_m (\dsim(f_i)  \sum_n b_n - c_{m, n} x_n) + \sum_n \dsin(f_i) x_n \ge \omega(f_i)$$

Or written with slack variables $s_m =  \sum_n b_n - c_{m, n} x_n$:

$$ y_m: \sum_n c_{m, n} x_n \ge b_m \forall m$$
$$ \lambda: \sum_m \dsim(f_i) s_m  + \sum_n \dsin(f_i) x_n \ge \omega(f_i)$$

The cut $\lambda$ in its last form is surprisingly short and that's where we aimed to go. 
The dual LP \ref{DualLPForCut} has at least an increase of the value of the optimal solution of $\omega(f_i)$.
So the same holds for its dual which is equivalent to last the inequalities.

\begin{Theorem}[Generation of branching cut]
The following inequality is true for all integer solutions:
$$ \lambda: \sum_m \dsim(f_i) s_m  + \sum_n \dsin(f_i) x_n \ge \omega(f_i)$$
The object increase by adding one cut of this kind is at least $\omega(f_i)$.
\end{Theorem}

\subsection{Thoughts about the new cuts}

First we apply this cut to the problem in \ref{Almost a real one} and we get:

$$ \frac{1}{3} s_1 + \frac{1}{12} s_2 + \frac{1}{12} s_3 + \frac{1}{3} s_4 + \frac{1}{12} s_5 \ge \frac{1}{6}$$
$$ \frac{9}{12} x_1 + \frac{6}{12} x_2 + \frac{6}{12} x_3 + \frac{6}{12} x_4 + \frac{6}{12} x_5 \le \frac{11}{12} - \frac{5}{6}$$
$$ \frac{3}{4} x_1 + \frac{1}{2} x_2 + \frac{1}{2} x_3 + \frac{1}{2} x_4 + \frac{1}{2} x_5 \le \frac{3}{4} $$

When applying $x_1 = 0$ as case in the original problem, you get the solution vector $(0 ,\frac{1}{2}, \frac{1}{4}, \frac{1}{4}, \frac{1}{2})$.
This solution is equalizing the above cut. And for the other case $x_1 = 1$ with solution vector $(1, 0, 0, 0, 0)$, this is also sharp.

Would be have chosen the file without normalization, we would have got:

$$x_1 + x_2 + x_3 + x_4 + x_5 \le \frac{3}{2}$$

At this cut $(0, \frac{1}{2}, \frac{1}{4}, \frac{1}{4}, \frac{1}{2})$ is equalizing the cut, but $(1, 0, 0, 0, 0)$ not. 
This could easily investigated more abstract.
Anyhow we state, normalization leads to sharper cuts, which is true in general.

\begin{remark}
The defined class of the cuts are sharp, in the sense that it can be used to get a proof that an integer solution is the optimal one.
\end{remark}

For binary problem this is not difficult to understand. 
Just make a case differentiation over all cases.
As we have a binary problem this is finite number, then this one derived cut is already sufficient.
In general you have to argue a bit cleverer, anyhow the remark is true.
\\
The above statement has no practical implication, as by making a case differentiation you already had a proof.
\\
Also the derived cut will be similar to the objective as above $x_1 + x_2 + x_3 + x_4 + x_5 \le \frac{3}{2}$ was already the objective function, but not with the real optimal integer objective value.
\\
\\
We will now get a little philosophical.
Consider you want to make a proof that an Integer solution with objective $\omega_0$ is an optimal one. 
By a big case differentiation you can produce one single cut, so that the best integer solution has to be almost $\omega_0$.
But the cut is already very similar to the objective function. 
So if you make a simple case differentiation after adding the cut, the objective will not increase in any branch at all.
Thus the big mighty cut is irrelevant for the proof at all. 
This suggest the below expectation:
\begin{remark}
  Many easy little steps are better than a few big complex steps.
\end{remark}

If you analyse the proof of the validity of the cut, things like parallelism of branching are not used at all.
This could lead to the wrong conclusion that the whole dual theory of concurrent branching is redundant.
The produced cuts in normal space yield the at least the objective increase as improvement with use of files at \ref{CentralAB} in the dual space.
This has not been shown explicitly here, but it is understood easily, when you change the starting model in \ref{DualLPForCut} so it uses more than one file.
\\
In the dual space you can calculate which files to use, measure the files and so the cuts. 
In dual space you just have better control of what you do.
\\
It would be only seeing the top of an iceberg, if the dual theory would have not been included here.
And last but not least the author first developed the dual theory for his idea of concurrent branching. 
Based on this idea he discovered, that it might be reapplied again to normal space.\\
\\
What's left to be done?\\
\\
The philosophical statement should be reasoned by some examples.\\
An implementation of the ideas to produce very quickly cuts or lookahead terms should be done, to really measure the usefulness of the theory.
Particular for binary problems it would be interesting to generate cuts on inequalities or equations with the discussed technique of fixing most of the dual variables.
\\
Furthermore it is most interesting to classify other cuts generation algorithm in our terms or vice versa. 
Also applying the sketched idea of effectiveness of lookahead terms and so cuts might to other cuts classes might be a fruitful idea.

\end{document}